\documentclass{article}
\usepackage{hiph-art}
\usepackage{graphicx}
\volnumber{19} \issuenumber{1} \edyear{2004}                             
\frompage{000} \topage{000}                                              
\recrevdate{November 2005}                                              
\def\rightmark{Visualizing Color Plasma Instabilities}
\def\leftmark{M. Strickland}
 
\title{Visualizing Color Plasma Instabilities} 
\authors{ 
{Michael Strickland$^1$%
\index{One, M.} 
}\\[2.812mm]
{\normalsize
\hspace*{-8pt}$^1$ Frankfurt Institute for Advanced Studies \\
Johann Wolfgang Goethe University \\
Max-von-Laue-Str. 1 \\
60438 Frankfurt am Main, Germany \\
}}
 
\abstract{
I discuss recent advances in the understanding of non-equilibrium gauge field 
dynamics in plasmas which have particle distributions which are locally 
anisotropic in momentum space.  In contrast to locally isotropic plasmas such 
anisotropic plasmas have a spectrum of soft unstable modes which are 
characterized by exponential growth of transverse (chromo)-magnetic fields at 
short times.  The long-time behavior of such instabilities depends on whether or 
not the gauge group is abelian or non-abelian.  I will report on recent 
numerical simulations which attempt to determine the long-time behavior of an 
anisotropic non-abelian plasma within hard-loop effective theory.  For novelty
I will present an interesting method for visualizing the time-dependence of SU(2) 
gauge field configurations produced during our numerical simulations.
}

 
\makeindex
\begin{document}
 
\maketitle

\section{Introduction}

One of the mysteries emerging from the RHIC ultrarelativistic heavy-ion 
collision experiments is that the matter produced in the collisions seems to be 
well-described by hydrodynamic models.  In order to apply hydrodynamical models 
the chief requirement is that the stress-energy tensor be isotropic in momentum 
space.  Additionally, current hydrodynamic codes also assume that they can use 
an equilibrium equation of state to describe the time evolution of the produced 
matter.  Therefore, the success of these models suggests that the bulk matter 
produced is {\em isotropic} and {\em thermal} at very early times, $t < 1$ fm/c. 
Estimates of the isotropization and thermalization times from perturbation 
theory \cite{BottomUp}, however, indicate that the time scale for thermalization 
is more on the order of $t \sim 2-3$ fm/c.  This contradiction has led some to 
conclude that perturbation theory should be abandoned and replaced by some other 
(as of yet unspecified) calculational framework. However, it has been proven 
recently that previous perturbative estimates of the isotropization and 
equilibration times had overlooked an important aspect of nonequilibrium gauge 
field dynamics, namely the possibility of {\em plasma instabilities}.

\begin{figure}[t]
\vspace{4mm}
\centerline{
\includegraphics[width=12.6cm]
{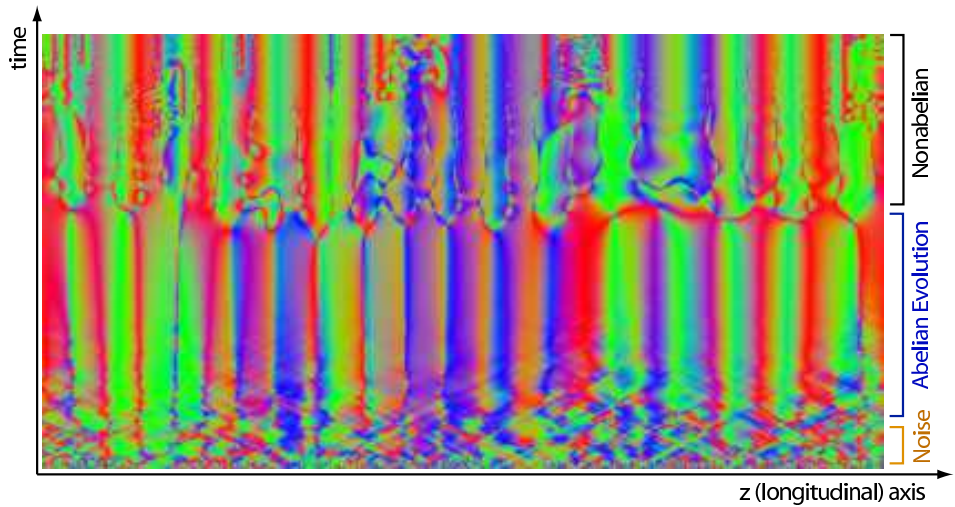}
}
\vspace{-2mm}
\caption{
Visualization of the spacetime dependence of the color-absolute value of the $x$-component 
of the induced current obtained from a 1+1-dimensional simulation.
\label{visualize_jtx_abs}}
\end{figure}

One of the chief obstacles to thermalization in ultrarelativistic heavy-ion 
collisions is the intrinsic expansion of the matter produced.  If the matter 
expands too quickly then there will not be time enough for its constituents to 
interact before flying apart into non-interacting particles and therefore the 
system will not reach thermal equilibrium.  In a heavy-ion collision the 
expansion which is most relevant is the longitudinal expansion of the matter 
since at early times it's much larger than the transverse expansion.  In the absence of 
interactions the longitudinal expansion causes the system to quickly become much 
colder in the longitudinal direction than in the transverse direction, 
$\langle p_L\rangle \ll \langle p_T\rangle$.  We can then ask how long it would take for 
interactions to restore isotropy in the $p_T$-$p_L$ plane.  In the bottom-up 
scenario \cite{BottomUp} isotropy is obtained by hard collisions between the 
high-momentum modes which interact via an isotropically screened gauge 
interaction.  The bottom-up scenario assumed that the underlying soft gauge 
modes responsible for the screening were the same in an anisotropic plasma as in 
an isotropic one. In fact, this turns out to be incorrect and in anisotropic 
plasmas the most important collective mode corresponds to an instability to 
transverse magnetic field fluctuations \cite{Stanislaw}.  Recent works have 
shown that the presence of these instabilities is generic for distributions 
which possess a momentum-space anisotropy \cite{CmodesPR,Arnold:2003rq} and have 
obtained the full hard-loop action in the presence of an anisotropy 
\cite{Mrowczynski:2004kv}.

Here I will discuss numerical results obtained within the last year which 
address the question of the long-time behavior of the instability evolution 
\cite{Arnold:2004ih,Rebhan:2004ur,Arnold:2005vb,Rebhan:2005re} within the hard-loop 
framework. This question is non-trivial in QCD due to the presence of non-linear 
interactions between the gauge degrees of freedom.  These non-linear 
interactions become important when the vector potential amplitudes become 
$\langle A\rangle_{\rm soft} \sim p_{\rm soft}/g \sim (g p_{\rm hard})/g$, where 
$p_{\rm hard}$ is the characteristic momentum of the hard particles.  In QED 
there is no such complication and the fields grow exponentially until 
$\langle A\rangle_{\rm hard} \sim p_{\rm hard}/g$ at which point the hard particles 
undergo large-angle scattering in the soft background field invalidating the 
assumptions underpinning the hard-loop effective action. Initial numerical toy 
models indicated that non-abelian theories in the presence of instabilities 
would ``abelianize'' and fields would saturate at  $\langle A\rangle_{\rm hard}$ 
\cite{Arnold:2004ih}.  This picture was largely confirmed by simulations of the 
full hard-loop gauge dynamics which assumed that the soft gauge fields depended 
only on the direction parallel to the anisotropy vector and time 
\cite{Rebhan:2004ur}.  However, recent numerical studies have now included the 
transverse dependence of the gauge field and it seems that the result is then 
that the gauge field's dynamics changes its behavior from exponential to linear 
growth when its amplitude reaches the soft scale, $\langle A\rangle_{\rm soft} \sim 
p_{\rm hard}$ \cite{Arnold:2005vb,Rebhan:2005re}.  This linear growth regime is 
characterized by a cascade of the energy pumped into the soft scale by the 
instability to higher momentum plasmon-like modes \cite{cascade}.  Below I will 
briefly describe the setup which is used by these numerical simulations and then 
discuss questions which remain in the study of non-abelian plasma instabilities.

In Sec.~\ref{visualization} I will present the method which is used to 
generate Fig.~\ref{visualize_jtx_abs} which shows the spacetime dependence of 
SU(2) gauge field configurations obtained from a 1+1 dimensional simulation of
an anisotropic quark-gluon plasma.  Due to the coarseness of our current 3+1-dimensional 
lattice simulations I will present only visualizations coming from 1d lattice simulations, 
however, the method is easily adapted to the 3+1-dimensional case.

\section{Discretized Hard-Loop Dynamics}%

At weak gauge coupling $g$, there is a separation of scales in hard momenta 
$|\mathbf p|=p^0$ of (ultrarelativistic) plasma constituents, and soft momenta 
$\sim g|\mathbf p|$ pertaining to collective dynamics. The effective field 
theory for the soft modes that is generated by integrating out the hard plasma 
modes at one-loop order and in the approximation that the amplitudes of the soft 
gauge fields obey $A_\mu \ll |\mathbf p|/g$ is that of gauge-covariant 
collisionless Boltzmann-Vlasov equations \cite{HTLreviews}. In equilibrium, the 
corresponding (nonlocal) effective action is the so-called hard-thermal-loop 
effective action which has a simple generalization to plasmas with anisotropic 
momentum distributions \cite{Mrowczynski:2004kv}. The resulting equations of 
motion are
\begin{eqnarray}
D_\nu(A) F^{\nu\mu} &=& -g^2 \int {d^3p\over(2\pi)^3} {1\over2|\mathbf p|} \,p^\mu\, 
{\partial f(\mathbf p) \over \partial p^\beta} W^\beta(x;\mathbf v) \, , \nonumber \\
F_{\mu\nu}(A) v^\nu &=& \left[ v \cdot D(A) \right] W_\mu(x;\mathbf v) \, , 
\label{eom}
\end{eqnarray}
where $f$ is a weighted sum of the quark and gluon distribution 
functions \cite{Mrowczynski:2004kv} and $v^\mu\equiv p^\mu/|\mathbf p|=(1,\mathbf v)$.

At the expense of introducing a continuous set of auxiliary fields 
$W_\beta(x;\mathbf v)$ the effective field equations are local.  These equations 
of motion are then discretized in spacetime and ${\mathbf v}$, and solved 
numerically in the temporal gauge, $A_0=0$.  The discretization in ${\mathbf v}$-space 
corresponds to including only a finite set of the auxiliary fields $W_\beta(x;\mathbf v_i)$ 
with $1 \leq i \leq N_W$. For details on the precise discretizations used see 
Refs.~\cite{Arnold:2005vb,Rebhan:2005re}.

\section{Numerical Results}

During the process of instability growth the soft gauge fields get the energy 
for their growth from the hard particles.  In an abelian plasma this energy 
grows exponentially until the energy in the soft field is of the same order of 
magnitude as the energy remaining in the hard particles.  As mentioned above in 
a non-abelian plasma one must rely on numerical simulations due to the presence 
of strong gauge field self-interactions.   Here I will present results for a particle
momentum distribution which has been squeezed along the $z$-axis.

\subsection{1+1 dimensional simulation}

\begin{figure}[t]
\vspace{4mm}
\centerline{
\includegraphics[width=9cm]
{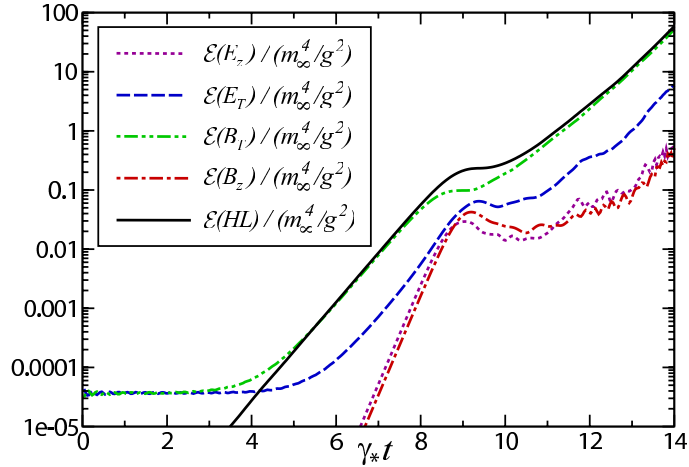}
}
\vspace{-2mm}
\caption{
Time dependence of energy densities obtained from 1+1-dimensional simulation.  Here
$\gamma^*$ is the growth rate for the dominant unstable mode and $N_w = 100$, see 
\cite{Rebhan:2004ur,Rebhan:2005re} for details.
\label{1dHLfig}}
\end{figure}

In Fig.~\ref{1dHLfig} I have plotted 
the time dependence of the energy density extracted from the hard particles obtained in 
a 1+1 dimensional simulation of an anisotropic plasma initialized with very weak 
random SU(2) color noise \cite{Rebhan:2004ur}.  From this Figure we see that after
an initial period during which various stable and unstable modes are competing
the system enters a period in which the energy density in transverse magnetic
fields, ${\cal E}(B_T)$ (green dot-dot-dashed), grows exponentially with a 
growth rate of the maximally unstable mode until $\gamma^*t \sim 9$.  In addition 
we see that the energy density extracted from the hard particles, ${\cal E}(HL)$ (solid black), 
also grows exponentially at the same rate during this time indicating that energy
extracted from the hard particles primarily goes into producing large amplitude
chromo-magnetic fields.  

At $\gamma^*t \sim 9$ the system enters the ``non-linear''
regime in which the three- and four-gluon couplings become relevant and there is
a brief slowdown in the growth of all quantities shown.  However, after some
field rearrangement the energy extracted from the particles and the produced soft 
fields then all grow at approximately the same rate.  In this case, the fields would continue
to grow until the energy in the soft fields was on the order of the energy in the
hard particles, at which time the hard particles would undergo large-angle deflections
off the soft fields.  This picture, however, only holds in QED or QCD restricted to
1+1 dimensional field configurations (fields are independent of directions transverse
to the instability vector).  As I will discuss in the next section when the dependence
of the chromo-electromagnetic fields on the transverse directions is included the
system's behavior changes dramatically in the non-linear regime.  However, in
Sec.~\ref{visualization} I will present visualizations which can be produced from
these (unrealistic and therefore somewhat academic) 1+1 dimensional simulations.

\subsection{3+1 dimensional simulation}

\begin{figure}[t]
\vspace{4mm}
\centerline{
\includegraphics[width=9cm]
{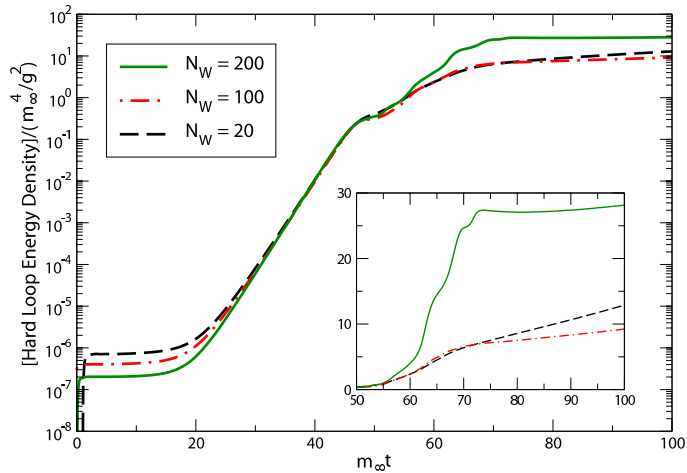}
}
\vspace{-2mm}
\caption{
Comparison of the energy transferred from hard to soft scales, $\mathcal 
E({\rm HL})$, for 3+1-dimensional simulations with $N_w=20,100,200$ 
on $96^3, 88^3, 69^3$ lattices. Inset shows late-time behavior on a linear 
scale.
\label{3dHLfig}}
\end{figure}

As mentioned in the previous section the late-time behavior of the system has a
dependence on the dimensionality of the fields assumed in the simulation.
In Fig.~\ref{3dHLfig} I have plotted the time dependence of the energy extracted 
from the hard particles obtained in a 3+1 dimensional simulation of an anisotropic 
plasma initialized with very weak random color noise \cite{Rebhan:2005re}.  As can 
be seen from this figure at $m_\infty t \sim 60$ there is a change from exponential 
to linear growth with the late-time linear slope decreasing as $N_W$ is increased.

The first conclusion that can be drawn from this result is that within non-abelian 
plasmas instabilities will be less efficient at isotropizing the plasma 
than in abelian plasmas.  However, from a theoretical perspective ``saturation'' 
at the soft scale implies that one can still apply the hard-loop effective 
theory self-consistently to understand the behavior of the system at late times. 

\section{Visualizations of 1+1 Results}\label{visualization}

In this section I present visualizations of the gauge-invariant currents generated from 
our 1+1 dimensional simulation.  As mentioned previously, these are merely of
academic interest since the 1+1 dimensional simulations do not seem to give the
correct late-time behavior for non-abelian gauge groups.  However, the visualizations
are still interesting in a pedagogical sense and allow one to easily ``see'' the
instabilities in action.

\begin{figure}[t]
\vspace{4mm}
\centerline{
\includegraphics[width=12.6cm]
{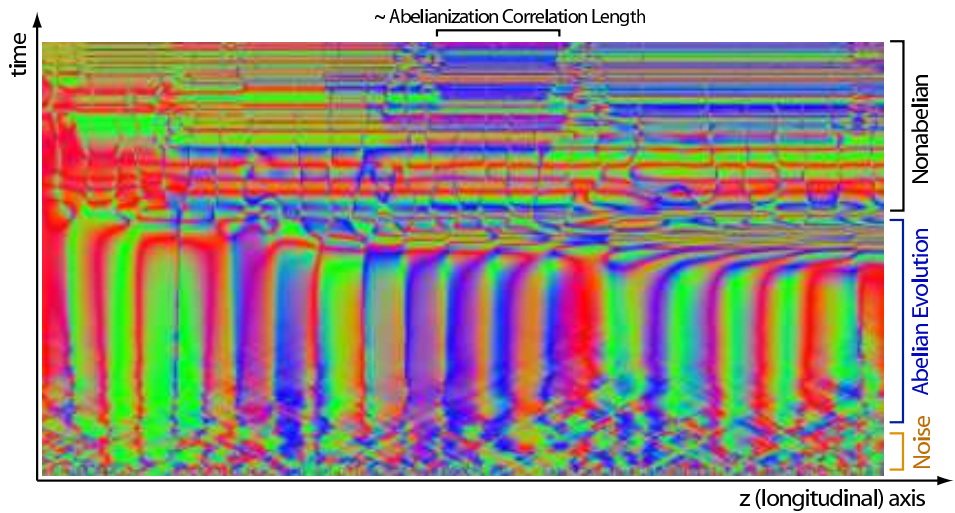}
}
\vspace{-2mm}
\caption{
Visualization of the spacetime dependence of the color-absolute value of the $x$-component 
of the induced current obtained from a 1+1-dimensional simulation.  Here all
color charges have been parallel transported to the leftmost site.
\label{visualize_jtx_pt_abs}}
\end{figure}

\begin{figure}[t]
\vspace{4mm}
\centerline{
\includegraphics[width=12.6cm]
{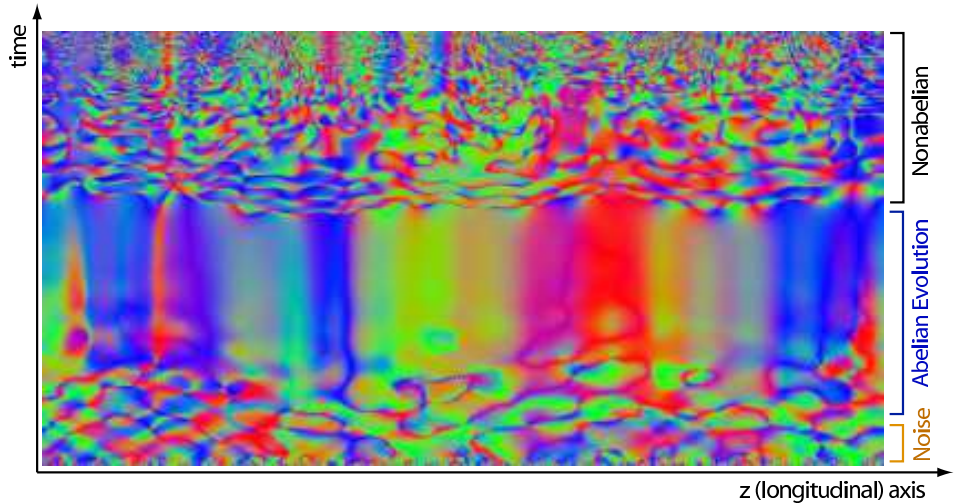}
}
\vspace{-2mm}
\caption{
Visualization of the spacetime dependence of the color-absolute value of the $z$-component 
of the induced current obtained from a 1+1-dimensional simulation.
\label{visualize_jz_abs}}
\end{figure}

To make these visualizations I use the representation of SU(2) vectors as \\ $O(3)/Z(2)$ 
vectors.  At each time (vertical axis in Figs.~\ref{visualize_jtx_abs}, \ref{visualize_jtx_pt_abs}, 
and \ref{visualize_jz_abs}) I loop over the one spatial dimension (horizontal axis), map each 
SU(2) vector to an O(3) vector, and normalize this vector such that it lies on a unit sphere 
centered at zero.  In order to remove the ambiguity coming from the antipodal equivalence associated 
with the Z(2) above I then take the absolute value of the vectors obtained from the initial map
so that the vectors all map to the positive octant.\footnote{The 8 to 1 map obtained by
taking the color-absolute value is overkill in the sense that some vectors which should not
be identified as equivalent are, however, it's the simplest way to implement the Z(2) invariance which
is not properly captured by a naive SU(2) to O(3) map.}  The vectors in this positive octant of the 
unit sphere are then mapped directly to RGB colors.

In Fig.~\ref{visualize_jtx_abs} I have plotted the spacetime dependence of the $x$-component
of the induced current, $j_x$, resulting from our 1+1 dimensional simulations with periodic
boundary conditions.  The bottom-most line in this figure contains the small-amplitude 
random color initial condition which was used.  Proceeding upwards in time the system first goes
through a noisy stage in which different unstable and stable modes are competing with the stable
left- and right-movers visible as diagonal lines resulting in a `criss-cross' pattern.  Next, the system 
goes through an abelian phase in which the current colors evolve almost independently with 
a typical spatial wavelength given by the wavelength of the maximally unstable mode 
\cite{Rebhan:2004ur,Rebhan:2005re}.  However, once the amplitudes of the color fields reach the
non-abelian scale the field self-interactions start to induce `splittings' and the currents begin
color-oscillating in time.\footnote{The spatial direction of the current is also changing in time but 
this cannot be gleaned from the visualizations presented here.  This results in some visually 
interesting artifacts which appear as `bubbly' vertical lines in Fig.~\ref{visualize_jtx_abs}.}

Since our simulations are performed in the temporal gauge, $A_0=0$, Fig.~\ref{visualize_jtx_abs} is the best way
to visualize the time-dependence of the gauge field since in this gauge the time-parallel transporter
is an identity matrix.  However, for assessing the spatial color correlations in the current it is
more appropriate to parallel transport the color matrices to a fixed spatial point for comparison.
In Fig.~\ref{visualize_jtx_pt_abs} I show the result of parallel transporting the same data shown in
Fig.~\ref{visualize_jtx_abs} to the left-most spatial lattice point ($z=0$).  As can be seen, in the 
abelian phase when the link variables are nearly unity, there is little difference between 
Figs.~\ref{visualize_jtx_abs} and \ref{visualize_jtx_pt_abs}; however, as the system approaches 
the non-abelian phase and the parallel transporters start to differ significantly from unity the
two figures are dramatically different.  In Fig.~\ref{visualize_jtx_pt_abs} at late times, in fact,
we can see that local regions where the color is approximately in the same direction emerge.  The
distance scale over which this occurs is the ``abelianization correlation length''.  This abelianization 
correlation length is finite and given approximately by the spatial wavelength of the most unstable 
mode.\footnote{For a more precise determination of the abelianization correlation length see 
Refs.~~\cite{Rebhan:2004ur,Rebhan:2005re}.}

Finally, I include, for comparison, a visualization of the current parallel to the anisotropy vector,
$j_z$, in Fig.~\ref{visualize_jz_abs}.  

\section{Conclusions}

The visualizations presented here are mostly novelty but they do allow one to easily
see the time evolution of the system and therefore can provide important intuitive information.
However, they can provide some insight into the evolution of the system at times
prior to when the cascade to higher-momentum modes would set in within the full 3+1-dimensional
dynamics.  Perhaps, when applied to finer 3+1-dimensional simulations and using surface finding 
routines they could even be used to better understand the complicated color dynamics occurring at 
late times.

Looking forward, I note that the latest 3+1-dimensional simulations 
\cite{Arnold:2005vb,Rebhan:2005re} have only presented results for distributions 
with a finite ${\mathcal O}(1-10)$ anisotropy and these seem to imply that in 
this case the induced instabilities will not have a significant effect on the hard 
particles.  This means, however, that due to the continued expansion of the 
system that the anisotropy will increase.  It is therefore important to 
understand the behavior of the system for more extreme anisotropies. 
Additionally, it would be very interesting to study the hard-loop dynamics in an 
expanding system.  Naively, one expects this to change the growth from 
$\exp(\tau)$ to $\exp(\sqrt\tau)$ at short times but there is no clear 
expectation of what will happen in the linear regime.  The short-time picture 
has been confirmed by early simulations of instability development in an 
expanding system of classical fields \cite{paulnew}.  It would therefore be 
interesting to incorporate expansion in collisionless Boltzmann-Vlasov transport 
in the hard-loop regime and study the late-time behavior in this case.

I note in closing that the application of this framework to phenomenologically 
interesting couplings is suspect since the results obtained strictly only apply 
at very weak couplings; however, the success of hard-thermal-loop perturbation 
theory at couplings as large as $g \sim 2$ \cite{HTLreviews,HTLpt} suggests that the 
nonequilibrium hard-loop theory might also apply at these large couplings.  For 
going to even larger couplings perhaps colored particle-in-cell simulations 
\cite{Dumitru:2005gp} could be used if they are extended to include collisions
and full 3+1 dynamics.

\section*{Acknowledgements}
I would like to thank A.~Rebhan and P.~Romatschke.
This work was supported by the Academy of Finland, contract no. 77744, and
the Frankfurt Institute for Advanced Studies.

\vfill\eject
\end{document}